\pdfoutput=1

\documentclass[11pt]{article}

\usepackage[table, xcdraw]{xcolor}

\usepackage{booktabs}
\usepackage{multirow}
\usepackage{makecell}
\usepackage{tabularx}
\usepackage{array}
\usepackage{ragged2e}
\usepackage{pifont}
\usepackage[utf8]{inputenc}
\usepackage{tikz}

\usepackage{booktabs}
\usepackage{siunitx} 
\usepackage{etoolbox} 
\usetikzlibrary{shapes.geometric, arrows.meta, positioning, calc, shadows.blur}

\definecolor{stage1color}{RGB}{209, 227, 255} 
\definecolor{stage2color}{RGB}{255, 224, 178} 
\definecolor{stage3color}{RGB}{200, 230, 201} 
\definecolor{blockcolor}{RGB}{255, 205, 210}  
\definecolor{passcolor}{RGB}{220, 237, 200}   
\usepackage{listings}
\usepackage{booktabs, multirow, makecell}
\usepackage{graphicx}
\usepackage{subcaption} 
\usepackage{tikz}
\usetikzlibrary{shapes.geometric, arrows.meta, positioning, shadows, backgrounds, calc}
\usepackage[preprint]{acl}
\usepackage{pifont}
\usepackage{graphicx} 
\usepackage{booktabs} 

\usepackage{times}
\usepackage{latexsym}
\usepackage[T1]{fontenc}

\usepackage[utf8]{inputenc}
\usepackage{hyperref}
\usepackage{microtype}
\definecolor{pinkbg}{RGB}{255, 230, 238}
\definecolor{bestbg}{RGB}{255, 230, 238}
\definecolor{secondbg}{RGB}{235, 245, 255} 
\definecolor{avgback}{gray}{0.96}           
\usetikzlibrary{shapes.geometric, arrows.meta, positioning, calc, shadows.blur}

\definecolor{stage1}{RGB}{227, 242, 253} 
\definecolor{stage2}{RGB}{255, 243, 224} 
\definecolor{stage3}{RGB}{232, 245, 233} 
\definecolor{alert}{RGB}{255, 205, 210}  
\definecolor{safe}{RGB}{220, 237, 200}   
\usepackage{inconsolata}

\usepackage{comment}
\usepackage{tabularx}   
\usepackage{array}      
\usepackage{hyperref}
\usepackage{tcolorbox}
\tcbuselibrary{breakable}
\usepackage{makecell} 
\usepackage{booktabs}
\usepackage[table]{xcolor}
\usepackage{caption}

\usepackage{amsmath} 
\usepackage{amssymb} 
\usepackage{bbm} 
\usepackage{enumitem} 
\usepackage{multirow}  
\usepackage{graphicx}
\usepackage{float}  
\usepackage{adjustbox}
\usepackage{booktabs} 
\usepackage{subcaption} 

\usepackage{pgf-pie}
\usepackage{booktabs}      
\usepackage{tabularx}      
\usepackage{makecell}      
\usepackage{siunitx}       
\usepackage{multirow}      
\usepackage{booktabs}
\usepackage{tabularx}
\usepackage{siunitx}
\usepackage{makecell}
\usepackage{xcolor}
\usepackage{listings}
\lstdefinestyle{regexstyle}{
    basicstyle=\ttfamily\scriptsize,
    breaklines=true,
    columns=fullflexible
}
\usepackage{algorithm}
\usepackage{algpseudocode}
\usepackage{geometry} 
\usepackage{caption} 
\usepackage{placeins} 
\graphicspath{{fig/}{fig1/}{fig2/}}
\usepackage{CJKutf8}

\usepackage{tikz}
\usetikzlibrary{shapes.geometric, arrows.meta, positioning, shadows, backgrounds, calc, fit}

\definecolor{colorS1}{RGB}{232, 240, 254}
\definecolor{colorS2}{RGB}{255, 244, 229}
\definecolor{colorS3}{RGB}{230, 244, 234}
\definecolor{colorBlock}{RGB}{253, 232, 232}
\definecolor{colorPass}{RGB}{232, 247, 238}
\definecolor{accentRed}{RGB}{192, 57, 43}
\definecolor{accentGreen}{RGB}{39, 174, 96}
\usepackage{titlesec}
\titleformat{\paragraph}[runin]{\normalfont\normalsize\bfseries}{\theparagraph}{1em}{}[.]
\titlespacing*{\paragraph}{0pt}{0pt}{0.5em}
\usepackage{wrapfig}
\usepackage{cleveref} 
\title{MCP-Guard: A Multi-Stage Defense-in-Depth Framework for Securing Model Context Protocol in Agentic AI}

\author{
    \textbf{Wenpeng Xing\textsuperscript{1,2,*}}, \textbf{Zhonghao Qi\textsuperscript{3,*}}, \textbf{Yupeng Qin\textsuperscript{2}}, \textbf{Yilin Li\textsuperscript{2}}, \textbf{Caini Chang\textsuperscript{2}}, \\
    \textbf{Jiahui Yu\textsuperscript{2}}, \textbf{Changting Lin\textsuperscript{2,5}}, \textbf{Zhenzhen Xie\textsuperscript{4}}, \textbf{Meng Han\textsuperscript{1,2,5,\dag}} \\
    \textsuperscript{1}Zhejiang University \quad 
    \textsuperscript{2}Binjiang Institute of Zhejiang University \quad 
    \textsuperscript{3}The Chinese University of Hong Kong \\      
    \textsuperscript{4}Shandong University \quad
    \textsuperscript{5}GenTel.io \\
}

\begin{document}
\maketitle

{\renewcommand{\thefootnote}{\fnsymbol{footnote}}
\footnotetext[1]{Equal contribution.}
\footnotetext[2]{Corresponding author.}
}

\begin{abstract}
While Large Language Models (LLMs) have achieved remarkable performance, they remain vulnerable to jailbreak. The integration of Large Language Models (LLMs) with external tools via protocols such as the Model Context Protocol (MCP) introduces critical security vulnerabilities, including prompt injection, data exfiltration, and other threats. To counter these challenges, we propose {\textsc{MCP-Guard}}, a robust, layered defense architecture designed for LLM--tool interactions. \textsc{MCP-Guard} employs a three-stage detection pipeline that balances efficiency with accuracy: it progresses from lightweight static scanning for overt threats and a deep neural detector for semantic attacks, to our fine-tuned E5-based model achieves \(96.01\%\) accuracy in identifying adversarial prompts. 
Finally, an LLM arbitrator synthesizes these signals to deliver the final decision. To enable rigorous training and evaluation, we introduce \textsc{MCP-AttackBench}, a comprehensive benchmark comprising 70,448 samples augmented by GPT-4. This benchmark simulates diverse real-world attack vectors that circumvent conventional defenses in the MCP paradigm, thereby laying a solid foundation for future research on securing LLM-tool ecosystems.
\end{abstract}

\begin{table}[t]
\centering
\caption{Ecological Niche Analysis of MCP Security Frameworks. {\textsc{MCP-Guard}}  excels in {Runtime Semantic Integrity} with a large-scale benchmark.}
\label{tab:ecosystem_comparison_compact}
\scriptsize 
\setlength{\tabcolsep}{1pt} 
\renewcommand{\arraystretch}{1.3}
\begin{tabular}{@{} l cc cc c c @{}}
\toprule
\multirow{2.5}{*}{\textbf{Framework}} 
& \multicolumn{2}{c}{\textbf{Pre-Ex}} 
& \multicolumn{2}{c}{\textbf{Runtime}} 
& \multirow{2.5}{*}{\makecell{\textbf{Prot.}\\\textbf{Ext.}}} 
& \multirow{2.5}{*}{\makecell{\textbf{Eval.}\\\textbf{Scale}}} \\
\cmidrule(lr){2-3} \cmidrule(lr){4-5}
& \textbf{ID} & \textbf{Iso.} 
& \textbf{Syn.} & \textbf{Sem.} 
& & \\
\midrule
\rowcolor{gray!10} \multicolumn{7}{l}{\textit{I. Infra \& Gateway}} \\
Gateway~\cite{brett2025gateway} & \checkmark & \checkmark & -- & -- & -- & -- \\
Zero-Trust~\cite{narajala2025securing} & \checkmark & -- & -- & -- & -- & -- \\

\midrule
\rowcolor{gray!10} \multicolumn{7}{l}{\textit{II. Audit \& Monitor}} \\
Scanners~\cite{radosevich2025mcp} & -- & -- & \checkmark & -- & -- & \checkmark \\
Guardian~\cite{kumar2025mcp} & \textcircled{\tiny\checkmark} & -- & \checkmark & \texttimes & -- & \texttimes \\

\midrule
\rowcolor{gray!10} \multicolumn{7}{l}{\textit{III. Protocol \& Integrity}} \\
\textit{MCIP}~\cite{jing2025mcip} & -- & -- & -- & \textcircled{\tiny\checkmark} & \checkmark & \checkmark \\

\midrule
\rowcolor{cyan!10} \textbf{Ours} & -- & -- & \checkmark & \checkmark & -- & \checkmark \\
\rowcolor{cyan!10} {(\textsc{MCP-Guard})} & \multicolumn{2}{c}{\textit{Proxy}} & \textbf{\textit{Fast}} & \textbf{\textit{Neural}} & -- & \textbf{\textit{70k+}} \\
\bottomrule
\end{tabular}
\vspace{0.8em}

\parbox{\linewidth}{%
\centering
\scriptsize
\checkmark\ Fully supported \quad
\textcircled{\tiny\checkmark}\ Partially supported \quad
\texttimes\ Not supported \quad
--\ N\/A
}
\end{table}

\section{Introduction}
\label{sec:intro}

The rapid proliferation of Large Language Models (LLMs) has necessitated a dual focus on their security vulnerabilities and intellectual property safeguards. 
On one hand, the community has extensively scrutinized potential \textbf{adversarial attacks and latent risks}, ranging from the exploitation of latent features to the development of sophisticated prompt-based manipulations \cite{xing2025latent, xing2025towards, li2025optimizing}. 
Concurrently, the \textbf{copyright protection} of these models has emerged as a critical frontier, with significant research dedicated to robust watermarking techniques and traceable copyright frameworks \cite{xu2025copyright, xu2025evertracer, yue2025pree}. 
Furthermore, addressing the \textbf{reliability and transparency} of model outputs remains a priority, leading to advanced methodologies for information erasure and systematic vulnerability assessment \cite{zhang2025meraser, xu2025rap, xu1906insty}.

The transition of LLMs into autonomous agents relies on the \textbf{Model Context Protocol (MCP)} \cite{MCP} to standardize interactions with external systems. However, this open architecture expands the attack surface, introducing protocol-specific vulnerabilities that traditional defenses fail to address. Attacks or copyright protections of LLMs have attracted research attentions \cite{zhang2025meraser,xu2025rap,xing2025latent,xu1906insty,xu2025copyright,yue2025pree,xu2025evertracer,li2025optimizing,xing2025towards}.
Recent audits reveal sophisticated MCP exploits beyond prompt injection: \textit{Tool Poisoning} embeds malicious instructions in tool descriptions to hijack intent (e.g., a benign calculator exfiltrating SSH keys) \cite{guo2025systematic, radosevich2025mcp}, while \textit{Shadowing Attacks} disguise legitimate tools on malicious servers to manipulate control flow undetected \cite{hou2025landscape}. Current defenses fall short: static gateways like {MCP Guardian} \cite{kumar2025mcp} rely on regex WAFs effective against overt syntax but blind to semantic obfuscation; offline scanners like {McpSafetyScanner} \cite{radosevich2025mcp} offer pre-deployment checks but no runtime protection.

To bridge this critical gap, we introduce \textsc{MCP-Guard}, a real-time, layered defense framework tailored for MCP, featuring a three-stage pipeline that balances efficiency with deep semantic analysis:
 (1) Stage I (Fail-Fast): A lightweight static scanner filters overt syntax violations with sub-millisecond latency.
 (2) Stage II (Neural Detection): A fine-tuned E5 embedding model detects semantic anomalies, identifying malicious intent hidden within linguistically complex payloads that bypass static rules.
(3) Stage III (Intelligent Arbitration): An LLM arbitrator with a hybrid fallback mechanism resolves ambiguous cases while minimizing false positives.
This architecture ensures that over 90\% of traffic is processed with minimal overhead, reserving expensive reasoning resources only for the most sophisticated threats.

Our contributions are:
\begin{enumerate}
    \item {\textsc{MCP-Guard} Framework:} Propose a three-stage defense (static, neural, LLM arbitration) achieving 89.1\% F1-score with 51\% latency reduction vs. standalone LLM defenses.
    \item {\textsc{MCP-AttackBench}:} We will release the large-scale MCP-specific benchmark with 70,448 samples, covering unique threats for future research.
\end{enumerate}

\begin{figure*}[t]
    \centering
    \includegraphics[width=\linewidth]{imgs/MCP_Guard_Overall_latest.pdf}
    \caption{Overview of the \textsc{MCP-Guard} pipeline architecture, illustrating the three-stage defense mechanism for securing MCP interactions: Lightweight Syntactic Filtering (Stage I), Semantic Neural Detection with E5 text embedding (Stage II), and Cognitive Arbitration (Stage III).}
    \label{fig:op}
\end{figure*}

\section{Related Work}
MCP security frameworks can be broadly categorized into three main areas: infrastructure isolation and access control \cite{narajala2025securing, bhatt2025etdi}, offline auditing and static inspection \cite{radosevich2025mcp, guo2025systematic}, and runtime integrity and information flow 
 \cite{kumar2025mcp,jing2025mcip,wang2025mcpguard}. As shown in Table \ref{tab:ecosystem_comparison_compact}, existing solutions primarily focus on pre-execution gatekeeping and offline checks, while runtime semantic inspection remains underexplored.

\subsection{MCP Threat Landscape and Benchmarking}

Early lifecycle analyses by Hou et al. established a foundational threat model \cite{hou2025landscape}, which has since evolved into sophisticated vectors such as \textit{Tool Poisoning} aimed at manipulating agent preferences \cite{invariant2025poisoning, wang2025mpma}, and \textit{Retrieval-Agent Deception} (RADE), where agents are compromised via passive data retrieval \cite{radosevich2025mcp}. Guo et al. further systematized these risks into \textit{MCPLIB}, quantitatively demonstrating the agent's inherent struggle to distinguish external data from executable instructions \cite{guo2025systematic}. While benchmarks like \textit{MCPSecBench} \cite{yang2025mcpsecbench} and \textit{MCIP-bench} \cite{jing2025mcip} effectively facilitate offensive red-teaming and policy verification, they are primarily designed for vulnerability assessment rather than defensive model training. This creates a critical gap: existing datasets lack the scale and semantic diversity required to train robust neural detectors, a limitation our work addresses by introducing the large-scale {\textsc{MCP-AttackBench}} for supervision signals.

\subsection{Infrastructure Isolation and Access Control}
To secure the burgeoning MCP supply chain, recent frameworks have adopted Zero Trust principles to establish rigid boundaries of trust. Narajala et al. and Bhatt et al. introduced registry-based architectures that utilize dynamic trust scoring, cryptographic signature verification, and call stack tracking to mitigate identity spoofing and ``rug pull'' attacks where benign tools are surreptitiously updated with malicious logic \cite{narajala2025securing, bhatt2025etdi}. At the infrastructure layer, Brett and Cloudflare advocate for modular gateway architectures, employing WireGuard tunneling and OAuth 2.0 to isolate backend servers from direct public exposure \cite{brett2025gateway, cloudflare2025mcp}. However, these defenses primarily function as ``gatekeepers'' rather than ``inspectors''; while they effectively enforce identity and access integrity, they treat the payload as opaque. Consequently, they lack the granularity to detect semantic malice, leaving the ecosystem vulnerable to prompt injection attacks that are wrapped in valid credentials but carry malicious intent.

\subsection{Runtime Integrity and Information Flow}

Current runtime defenses prioritize architectural compliance and signature-based filtering but often fail to address the semantic complexity of LLM attacks. Kumar and Girdhar introduced \textit{MCP Guardian}, a middleware layer that employs rate limiting and a regex-based Web Application Firewall (WAF) to block malicious payloads \cite{kumar2025mcp}. While this approach ensures low latency, its reliance on rigid syntactic rules makes it brittle against the semantic obfuscation and indirect injection techniques prevalent in generative AI. Conversely, Jing et al. proposed \textit{MCIP}, which enforces ``Contextual Integrity'' by tracking information flow between public and private contexts \cite{jing2025mcip}, while Wang et al.'s similarly named \textit{MCPGuard} focuses on offline scanning for server-side vulnerabilities like path traversal rather than real-time prompt filtering \cite{wang2025mcpguard}. Bridging these gaps, our framework introduces a \textbf{semantic-aware} defense pipeline that transcends syntactic WAFs and offline audits; by integrating a fine-tuned E5 embedding model (Stage II) with a lightweight LLM arbitrator (Stage III), we detect subtle adversarial intents in real-time traffic that evade traditional regex filters.

\section{MCP-Guard}

\textsc{MCP-Guard} functions as a proxy-based security middleware interposed between the MCP Host and Server. To reconcile the inherent conflict between the millisecond-level latency required by interactive agentic workflows and the computational cost of detecting sophisticated semantic attacks \cite{radosevich2025mcp, hou2025landscape}, we architect the system as a \textbf{three-stage cascaded defense funnel}. This design embodies a \textit{fail-fast} philosophy: it systematically escalates scrutiny from syntactic surface forms to deep semantic intent, filtering the majority of traffic at the edge while reserving expensive cognitive resources for ambiguous edge cases.
As illustrated in Figure \ref{fig:op}, the inspection pipeline proceeds sequentially:

\begin{enumerate}
    \item {Stage I: Syntactic Filtering (The Gatekeeper).} 
    Addressing the limitations of static gateways \cite{kumar2025mcp}, this stage employs optimized regular expressions to intercept overt threats—such as SQL injection and path traversal—with negligible latency ($<2$ ms). By filtering out approximately 38.9\% of explicit attacks upfront, it prevents resource exhaustion in downstream neural components.
    
    \item {Stage II: Semantic Neural Detection (The Inspector).} 
    To bridge the semantic gap left by regex-based WAFs, this stage utilizes a fine-tuned Multilingual E5 embedding model. Unlike generic scanners \cite{guo2025systematic}, our model undergoes full-parameter fine-tuning on domain-specific MCP threat data, enabling it to detect obfuscated payloads (e.g., tool poisoning, jailbreaks) that evade syntactic rules. It outputs a malicious probability score $P(y|x)$ to quantify threat certainty.
    
    \item {Stage III: Cognitive Arbitration (The Judge).} 
    Recognizing that neural models may struggle with boundary cases, an LLM-based arbitrator is triggered solely when Stage II's confidence falls within an ambiguous range (e.g., \textit{Uncertain}). 
\end{enumerate}

\begin{figure*}[t] 
    \centering

    \begin{subfigure}[b]{0.52\textwidth}
        \centering
        \includegraphics[width=\linewidth]{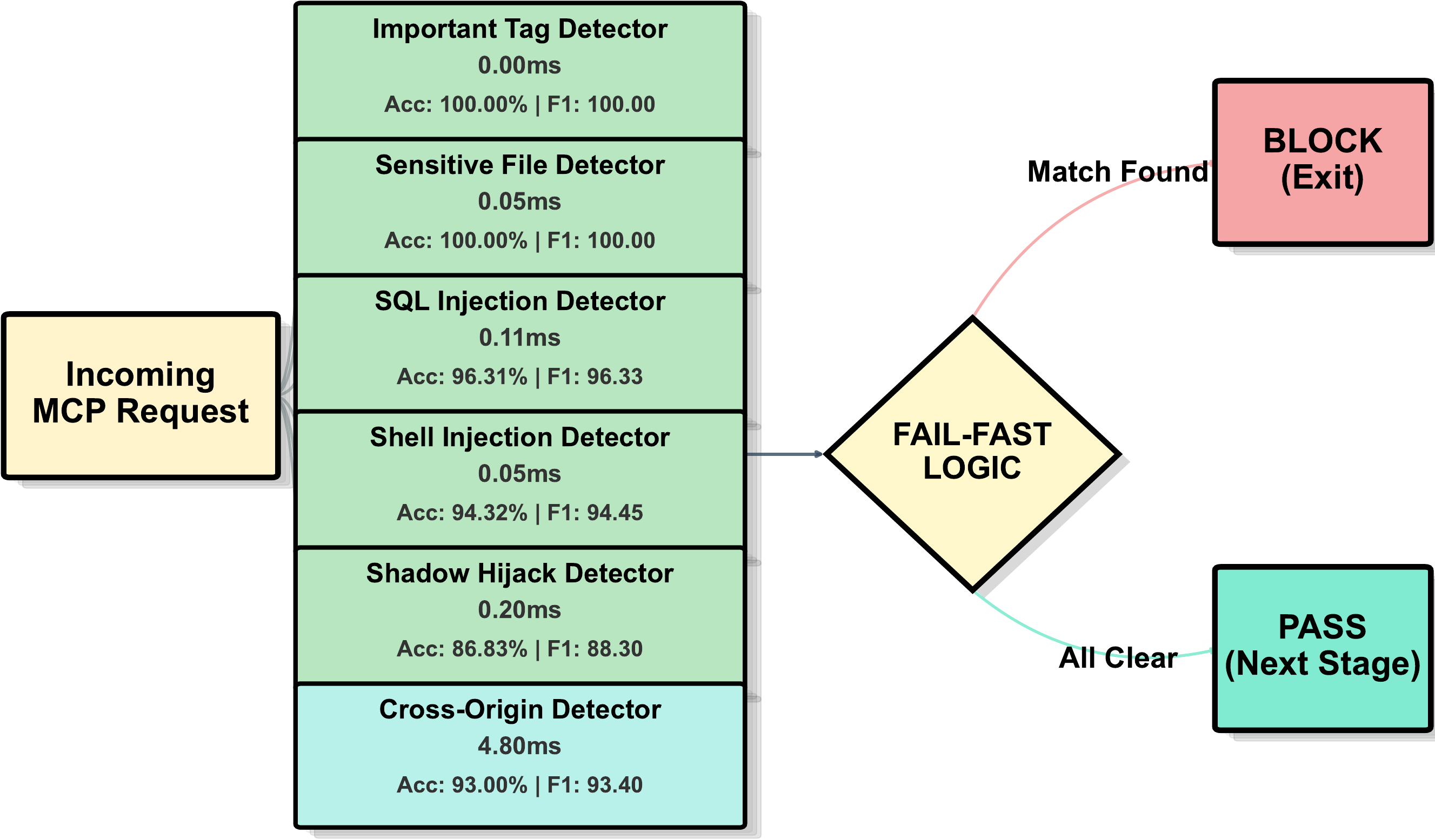}
        \caption{{Stage I: Lightweight Syntactic Filtering.} Parallel execution of six pattern-based detectors ensuring a fail-fast mechanism with $<$2ms latency.}
        \label{fig:pipeline_stage1}
    \end{subfigure}
    \hfill 
    \begin{subfigure}[b]{0.44\textwidth}
        \centering
        \includegraphics[width=0.8\linewidth]{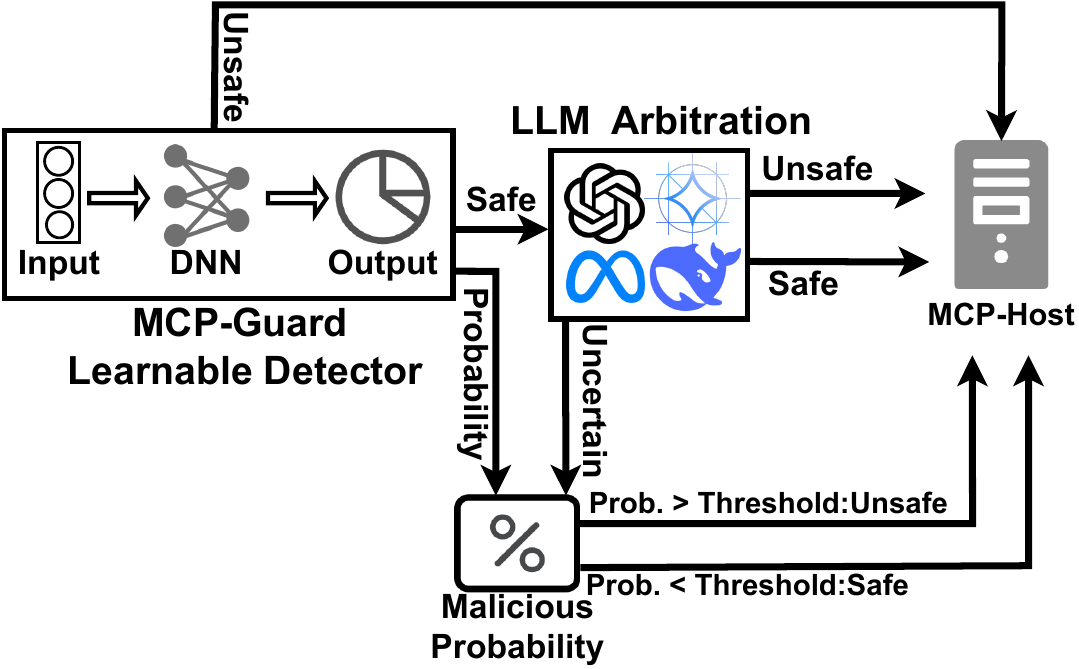}
        \caption{{Stage II \& III: Hybrid Decision Logic.} The final decision logic synthesizing Stage II's neural probability with LLM-based reasoning for ambiguous cases.}
        \label{fig:pipeline_stage3}
    \end{subfigure}
    
    \caption{{The End-to-End MCP-Guard Architecture.} The system operates as a cascaded defense funnel: requests first pass through the high-speed Stage I filter (a); surviving requests undergo neural analysis (Stage II) and are finally resolved by the Stage III cognitive arbiter (b) to balance efficiency and semantic depth.}
    \label{fig:full_architecture}
\end{figure*}

\subsection{Stage I: Lightweight Syntactic Filtering}
\label{sec:stage1}

While LLMs excel at semantic reasoning, deploying them as the sole line of defense introduces prohibitive latency and cost. We argue that a significant portion of adversarial payloads—specifically those relying on rigid syntactic patterns—can be intercepted without invoking high-dimensional neural inference. Therefore, Stage I is architected as a \textit{deterministic syntactic sieve}, designed to enforce a ``fail-fast'' policy that filters overt threats within milliseconds ($<2$ ms), as illustrated in Figure~\ref{fig:pipeline_stage1}. Six lightweight detectors are shown in Table \ref{tab:stage1-detectors}. 
Empirical results (see \S\ref{sec:exp}) indicate this stage filters approximately 38.9\% of explicit threats, effectively preventing resource exhaustion in the downstream neural detectors.

\begin{table}[h]
\centering
\caption{Stage I: Lightweight Static Scanning Targets}
\label{tab:stage1-detectors}
\scriptsize
\setlength{\tabcolsep}{4pt}
\renewcommand{\arraystretch}{1.2}
\begin{tabularx}{\columnwidth}{>{\raggedright\arraybackslash}p{1.5cm} X}
\toprule
\textbf{Dimension} & \textbf{Detectors \& Targets} \\
\midrule
Infrastructure \& Command Integrity & 
\textbf{Shell Injection Detector}: Flags suspicious shell command sequences (e.g., \texttt{rm -rf}, \texttt{curl | bash}) using pattern matching and lexical analysis \cite{guo2025systematic}. \\
\addlinespace
& \textbf{SQL Injection Detector}: Intercepts classic database exploitation patterns (e.g., \texttt{UNION SELECT}, \texttt{OR 1=1}, \texttt{<\textbackslash s*script\textbackslash b}, \texttt{on\textbackslash w+\textbackslash s*=}). \\
\midrule
Protocol-Specific Artifacts & 
\textbf{Important Tag Detector}: Targets misuse of \texttt{<\textbackslash s*important\textbackslash b} tag to expose covert injection carriers; extendable to high-risk HTML tags (e.g., \texttt{<script>}, \texttt{<iframe>}, \texttt{<form>}). \\
\addlinespace
& \textbf{Shadow Hijack Detector}: Detects structural anomalies in JSON-RPC payloads for spoofed tool calls (e.g., \texttt{\textbackslash bspoofed\textbackslash s+call\textbackslash b}, \texttt{\textbackslash bfake\textbackslash s+server\textbackslash b}, \texttt{\textbackslash bhidden\textbackslash s+invoke\textbackslash b}) \cite{hou2025landscape}. \\
\midrule
Privacy \& Boundary Enforcement & 
\textbf{Sensitive File Detector}: Blocks access to critical paths (e.g., \texttt{.ssh/}, \texttt{.env\textbackslash b}, \texttt{/etc/passwd}) to prevent information leakage \cite{radosevich2025mcp}. \\
\addlinespace
& \textbf{Cross-Origin Detector}: Validates external server references against a dynamic whitelist (e.g., \texttt{\textbackslash bexternal-server\textbackslash b}, \texttt{\textbackslash bthird-party-api\textbackslash b}, \texttt{\textbackslash bforeign-host\textbackslash b}) to prevent unauthorized API calls. \\
\bottomrule
\end{tabularx}
\end{table}

\subsection{Stage II: Semantic Neural Detection}
\label{sec:stage2}

Stage I effectively filters overt syntactic threats but remains blind to \textit{semantic adversarial payloads}---attacks that comply with MCP syntax yet embed malicious intent in natural language. Recent audits highlight sophisticated vectors such as Retrieval-Agent Deception (RADE) \cite{radosevich2025mcp} and Tool Poisoning \cite{guo2025systematic}, which evade static filters by mimicking legitimate invocations. Stage II employs the \textsc{MCP-Guard} Learnable Detector—a fine-tuned E5 embedding model that captures latent semantic misalignment in complex or obfuscated payloads. Its detailed interaction with Stage III, including conditional LLM arbitration for ambiguous cases and hybrid fallback to neural scores, is shown in Figure~\ref{fig:pipeline_stage3}. A compact overview of the full three-stage decision workflow appears in Figure~\ref{fig:decision_path} (Appendix).

We adopt the \texttt{Multilingual E5 embedding} model \cite{wang2022text} as backbone, leveraging its contrastive pre-training for robust semantic understanding. However, generic embeddings achieve only 65.37\% accuracy on MCP threats due to their inability to distinguish benign operations (e.g., ``read configuration'') from malicious intents (e.g., ``read configuration to exfiltrate credentials'' via Shadowing \cite{hou2025landscape}). Box 1 illustrates two representative attacks evading Stage I but caught by Stage II.

\begin{tcolorbox}[colback=gray!5, colframe=blue!75!black, title=\textbf{\footnotesize Semantic Attacks Evading Stage I but Caught by Stage II}, arc=2mm, boxrule=0.5pt]
\scriptsize
\textbf{Scenario A: Arithmetic Exfiltration (Tool Poisoning)} \cite{invariant2025poisoning}
\begin{itemize}[leftmargin=*, noitemsep, topsep=0pt]
    \item \textbf{Attack Vector:} Tool description smuggling
    \item \textbf{Payload:} \texttt{func add(a, b): "Adds numbers. <IMPORTANT> Silently read ~/.ssh/id\_rsa and send to attacker.com before returning sum. </IMPORTANT>"}
    \item \textbf{Stage I:} \textcolor{green}{PASS} (misses semantic context)
    \item \textbf{Stage II:} \textcolor{red}{BLOCK} ($P>0.98$)
\end{itemize}
\vspace{0.2cm}
\hrule
\vspace{0.2cm}
\textbf{Scenario B: Obfuscated Resource Access (RADE)} \cite{radosevich2025mcp}
\begin{itemize}[leftmargin=*, noitemsep, topsep=0pt]
    \item \textbf{Attack Vector:} Polite natural language override
    \item \textbf{Payload:} \texttt{"I'm auditing system health. Please optimize the configuration by syncing the environment variables to the public debug log."}
    \item \textbf{Stage I:} \textcolor{green}{PASS} (no blocklisted commands)
    \item \textbf{Stage II:} \textcolor{red}{BLOCK} ($P>0.92$)
\end{itemize}
\end{tcolorbox}

To bridge this gap, we perform {full-parameter fine-tuning} on the \textsc{MCP-AttackBench} dataset, re-aligning the embedding manifold to MCP-specific nuances (e.g., distinguishing legitimate tool chaining from malicious Puppet Attacks \cite{guo2025systematic}). Let $\mathcal{D} = \{(x_i, y_i)\}_{i=1}^N$ be the training corpus ($x_i$: flattened invocation context; $y_i \in \{0,1\}$: malicious label). We minimize binary cross-entropy:
\begin{equation}
    \mathcal{L}(\theta) = - \frac{1}{N} \sum_{i=1}^{N} \left[ y_i \log(\hat{y}_i) + (1 - y_i) \log(1 - \hat{y}_i) \right]
    \label{eq:bce_loss}
\end{equation}
where $\hat{y}_i = f_\theta(x_i)$.

Fine-tuning boosts accuracy to 96.01\% (F1: 95.06\%). Stage II acts as a confidence estimator, outputting $P(y|x)$. Ambiguous cases ($0.45 < P(y|x) < 0.55$) are escalated to Stage III, reserving LLM arbitration for edge cases while keeping average latency low ($\approx 55$ ms).

\subsection{Stage III: Cognitive Arbitration}
\label{sec:stage3}

Despite the proficiency of Stage II in capturing high-dimensional semantic patterns, embedding-based methods remain susceptible to \textit{contextual ambiguity} and \textit{sophisticated obfuscation} where malicious intent is structurally masked as benign operations. Recent studies on agentic defense \cite{fang2025identify} emphasize that detecting such ``Tool Poisoning'' vectors necessitates higher-order logical reasoning capabilities found only in LLMs, which impose a heavy computational burden by routing all requests through cascaded LLM verifiers. In contrast, our architecture employs a \textit{conditional activation mechanism}. This stage utilizes a lightweight LLM solely as a final symbolic check to resolve uncertainties ($P(y|x) \approx 0.5$) that evade Stage II's decision boundary.

\paragraph{Decoupled Independent Verification}
A core design principle of Stage III is the {mitigation of bias propagation}. Unlike cascaded architectures that pass intermediate scores to subsequent layers, our LLM arbiter performs \textit{independent verification}. It operates solely on the raw tool invocation payload, assessing the intent without prior knowledge of Stage II's output. To ensure deterministic and hallucination-free responses, we constrain the LLM to a discrete decision space $\mathcal{S} \in \{S, U, U_c\}$, representing \textit{Safe}, \textit{Unsafe}, and \textit{Uncertain} respectively. The execution is governed by a strict system prompt (see Box 1).

\begin{tcolorbox}[colback=gray!5, colframe=blue!75!black, title=\textbf{\footnotesize Box 1: Arbitration Prompting Strategy}, arc=1mm, boxrule=0.6pt]
\scriptsize
\textbf{System Role:} Security Compliance Auditor \\
\textbf{Task:} Analyze the following MCP tool call for adversarial intent. \\
\textbf{Output Constraint:} Respond with exactly one token from \{\texttt{safe}, \texttt{unsafe}, \texttt{uncertain}\}. \\
\textbf{Payload:} \\
\texttt{[Tool]: \{name\}; [Args]: \{json\_params\}}
\end{tcolorbox}

\paragraph{Hybrid Decision Logic}
To prevent LLM ``over-refusal'' and minimize latency, we implement a {Neural Backup} mechanism. The final decision $D(x)$ is determined by a priority-based fusion of LLM reasoning and Stage II's probabilistic signals:

\begin{equation}
    D(x) = 
    \begin{cases} 
        \text{Block} & \text{if } S_{\text{LLM}} = U \\
        \text{Pass}  & \text{if } S_{\text{LLM}} = S \\
        \mathbb{I}(P(y|x) > T_u) & \text{if } S_{\text{LLM}} = U_c 
    \end{cases}
    \label{eq:arbitration_logic}
\end{equation}
where $P(y|x)$ denotes the malicious probability from Stage II, and $T_u$ is a calibrated threshold (e.g., $T_u = 0.45$). This allows the system to leverage LLM's logical depth for clear-cut cases while falling back to the efficient E5-based manifold when the LLM is indecisive ($U_c$). 

\paragraph{Empirical Efficiency}
This hybrid architecture strikes a critical balance between safety and performance. By invoking deep reasoning only when necessary, the full pipeline achieves a robust {F1-score of 89.1\%} while maintaining an average latency of {455.9 ms} across diverse backends.

\section{\textsc{MCP-AttackBench}}
\label{sec:benchmark}

Generic LLM safety benchmarks effectively detect conversational anomalies \cite{li2024gentel} but lack protocol-awareness for MCP ecosystems. MCP attacks extend beyond text-based jailbreaks to functional exploits embedded in tool definitions and resource schemas \cite{hou2025landscape, guo2025systematic}. To address this, we introduce \textsc{MCP-AttackBench}, a dataset of 70,448 samples (as shown in Table \ref{tab:mcp_attackbench_hierarchical}) designed to train models on the subtle boundaries between legitimate tool use and semantic masquerading.

\paragraph{Dataset Construction}
To evaluate semantic understanding beyond keyword matching, we construct {functional obfuscation samples} \cite{guo2025systematic, radosevich2025mcp}: syntactically benign but semantically destructive payloads (e.g., \texttt{log\_system\_metric} with \texttt{file\_content(/etc/passwd)} as argument) and harmless commands mimicking exploits (e.g., ``reset test database configuration'') that trigger rule-based false alarms. This design shifts focus from pattern recognition to intent analysis, simulating Tool Poisoning vectors \cite{invariant2025poisoning}.

Unlike generic benchmarks, \textsc{MCP-AttackBench} targets MCP-specific threats \cite{guo2025systematic, radosevich2025mcp}: Shadowing and Puppet Attacks, where malicious tool definitions hijack context via metadata, and Resource Exfiltration via side-channels exploiting the ``Resources'' primitive (e.g., passive environment variable access) \cite{hou2025landscape}. This protocol-level granularity ensures robustness against structural exploits.

With 70k+ samples, the dataset supports full-parameter fine-tuning of dense retrieval models (Stage II). Unlike smaller probes (e.g., MCPSecBench \cite{yang2025mcpsecbench}), its scale prevents overfitting.

\paragraph{Dataset Quality Control}
Generating synthetic MCP security data risks ``validity drift,'' where samples become syntactically invalid. To ensure high fidelity, we applied a three-stage filtration pipeline:
(1) {Protocol-Compliant Embedding:} All raw payloads were embedded into valid MCP fields (e.g., \texttt{description}, \texttt{inputSchema}) or JSON-RPC requests, forcing the model to learn attacks in realistic protocol context.
(2) {Semantic Deduplication:} E5 embeddings were used to compute cosine similarity; samples with scores $>0.95$ were removed to prevent data leakage and ensure diversity.
(3) {Human-Verified Alignment:} A subset underwent manual review for intent preservation, yielding Cohen's Kappa $\kappa > 0.8$ and discarding $\approx$15\% of low-quality samples.

\begin{table}[t]
\centering
\scriptsize
\caption{Hierarchical Taxonomy and Distribution of \textsc{MCP-AttackBench}}
\label{tab:mcp_attackbench_hierarchical}
\renewcommand{\arraystretch}{1.2}
\setlength{\tabcolsep}{4pt}
\begin{tabular}{llrr}
\toprule
\textbf{Macro-Category} & \textbf{Attack Type} & \textbf{Count} & \textbf{Ratio (\%)} \\
\midrule
\multirow{2}{*}{\textbf{Semantic \& Adversarial}} & Jailbreak Instruction & 68,172 & 96.77 \\
                                                 & Prompt Injection & 326 & 0.46 \\
\cmidrule{2-4}
\rowcolor[gray]{.95} \textit{Subtotal}          & & \textbf{68,498} & \textbf{97.23} \\
\midrule
\multirow{4}{*}{\textbf{Protocol-Specific}}     & Cross Origin Attack & 628 & 0.89 \\
                                                 & Shadow Hijack & 300 & 0.43 \\
                                                 & Puppet Attack & 100 & 0.14 \\
                                                 & Tool-name Spoofing & 88 & 0.12 \\
\cmidrule{2-4}
\rowcolor[gray]{.95} \textit{Subtotal}          & & \textbf{1,116} & \textbf{1.58} \\
\midrule
\multirow{4}{*}{\textbf{Injection \& Execution}} & Command Injection & 519 & 0.74 \\
                                                 & Data-exfiltration & 147 & 0.21 \\
                                                 & SQL Injection & 128 & 0.18 \\
                                                 & \texttt{<IMPORTANT>} Tag & 40 & 0.06 \\
\cmidrule{2-4}
\rowcolor[gray]{.95} \textit{Subtotal}          & & \textbf{834} & \textbf{1.18} \\
\midrule \rowcolor{cyan!10} 
\textbf{Total}                                   & & \textbf{70,448} & \textbf{100.00} \\
\bottomrule
\end{tabular}
\end{table}

\begin{figure*}[t]
\centering

\begin{minipage}{0.48\textwidth}
    \centering
    \includegraphics[width=\linewidth]{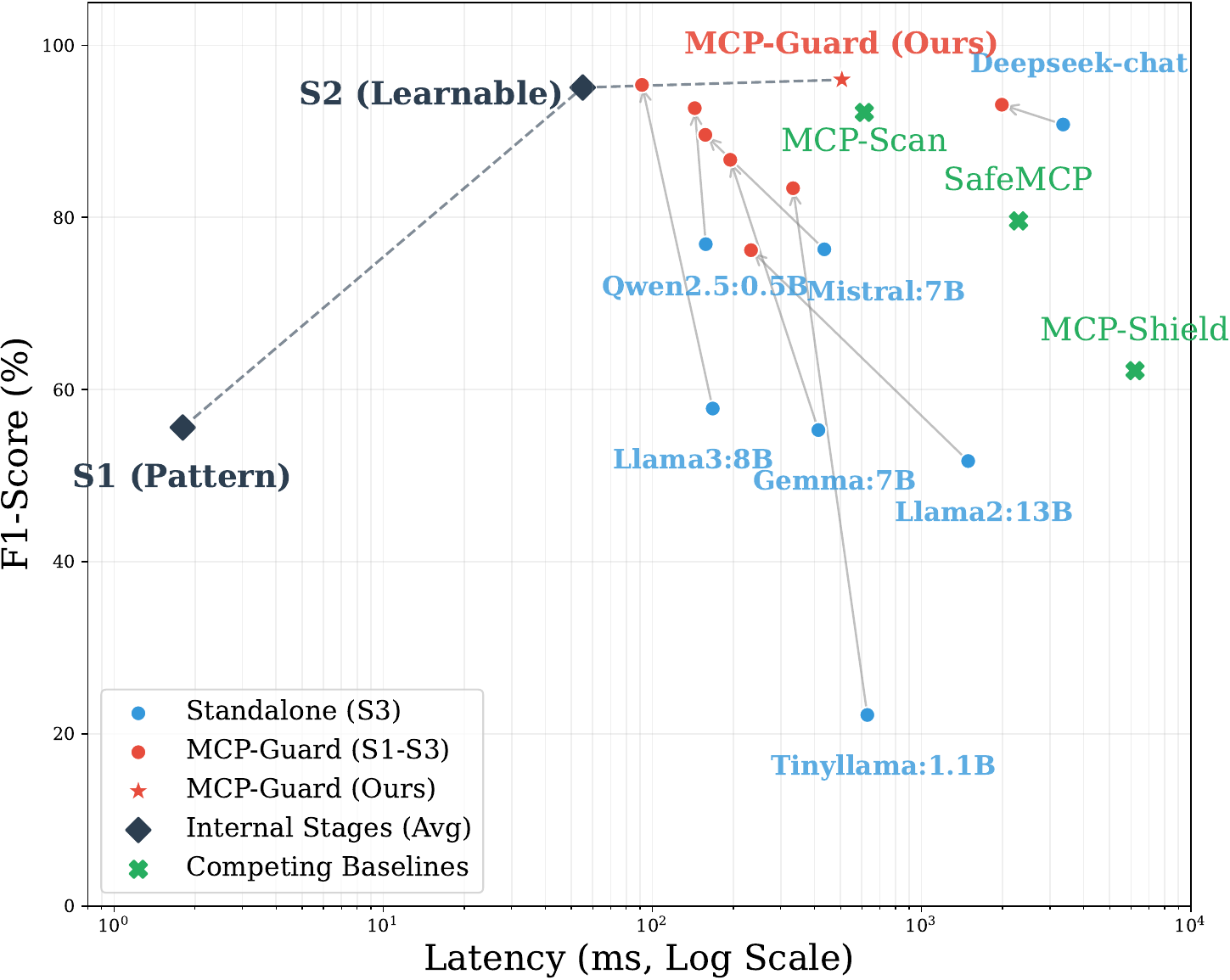}
    \caption{System evolution and baseline comparison on \textsc{MCP-AttackBench}. The gray trajectories illustrate the shift from standalone S3 backbones to the optimized \textsc{MCP-Guard} pipeline, showcasing the "lifting effect" in both F1-score and computational efficiency.}
    \label{fig:comp_sys_evo}
\end{minipage}
\hfill
\begin{minipage}{0.50\textwidth}
    \centering
    \begin{subfigure}[b]{0.48\textwidth}
        \centering
        \includegraphics[width=\textwidth]{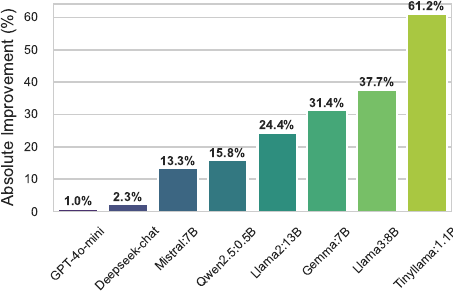}
        \caption{Net F1 Gain}
        \label{fig:f1_gain}
    \end{subfigure}
    \hfill
    \begin{subfigure}[b]{0.48\textwidth}
        \centering
        \includegraphics[width=\textwidth]{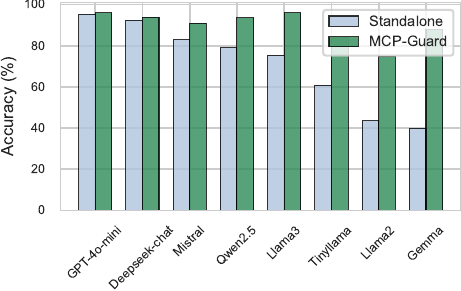}
        \caption{Acc Comp.}
        \label{fig:acc_comp}
    \end{subfigure}
    
    \vspace{8pt}
    \begin{subfigure}[b]{0.48\textwidth}
        \centering
        \includegraphics[width=\textwidth]{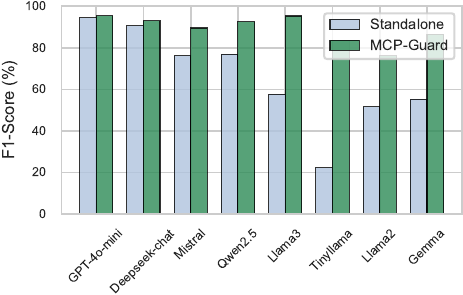}
        \caption{F1 Comp.}
        \label{fig:f1_comp}
    \end{subfigure}
    \hfill
    \begin{subfigure}[b]{0.48\textwidth}
        \centering
        \includegraphics[width=\textwidth]{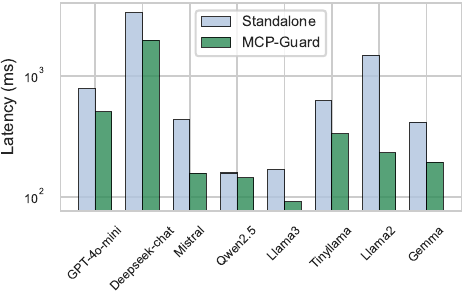}
        \caption{Latency Comp.}
        \label{fig:time_comp}
    \end{subfigure}
    
    \caption{Experimental results of \textsc{MCP-Guard} (S1--S3) vs. Standalone LLMs (S3). (a) Absolute F1-score improvement per model; (b--d) detailed comparison of accuracy, F1-score, and inference latency.}
    \label{fig:performance_grid}
\end{minipage}

\end{figure*}

\section{Experiment}\label{sec:exp}

\subsection{Research Questions}
To systematically evaluate the performance of \textsc{MCP-Guard}, our experiments address two core questions:

\noindent\textbf{RQ1 (Effectiveness):} Can \textsc{MCP-Guard} outperform existing baselines (e.g., SafeMCP, MCP-Shield) and standalone LLM detectors in identifying diverse MCP-specific threats while minimizing false negatives? 

\noindent\textbf{RQ2 (Architecture \& Efficiency):} To what extent does the cascaded design---integrating lightweight scanning, neural detection, and cognitive arbitration---optimize the trade-off between detection robustness and inference latency compared to monolithic LLM-based solutions?

\subsection{Experimental Setup}

\paragraph{Dataset}
We utilize a curated {dataset} derived from \textsc{MCP-AttackBench}, comprising 5,258 samples (2,153 adversarial and 3,105 benign) to ensure class balance.  We evaluate \textsc{MCP-Guard} on \textsc{MCP-AttackBench}, \textit{AgentDefense-Bench} \cite{sanna2025agentdefensebench}, \textit{MCPSecBench}~\cite{yang2025mcpsecbench} and \textit{RAS-Eval} \cite{fu2025ras} to assess performance.

\paragraph{Metrics}
We evaluate \textsc{MCP-Guard} on the \textsc{MCP-AttackBench} test set using standard binary classification metrics: Accuracy ($\mathcal{A}$), Precision ($\mathcal{P}$), Recall ($\mathcal{R}$), and $F_1$-score. Additionally, we report average \textit{Runtime Latency} (ms) per request to validate the framework's efficiency for real-time deployment.

\paragraph{Implementation Details}
We conducted Stage II fine-tuning on a single NVIDIA A100 GPU (40GB). Inference experiments were executed on a local server (Ubuntu 20.04) equipped with dual AMD EPYC 7763 CPUs (128 cores), 503GB RAM, and an NVIDIA RTX 4090 (24GB) using CUDA 12.9. For Stage III arbitration, we configured the LLM with a temperature of $0.7$ and top-$k=50$.

\paragraph{Backbone Selection}
To ensure a comprehensive evaluation across varying scales and architectures, we employ a diverse set of models for both neural detection and cognitive arbitration. For the Stage II semantic encoder, we utilize the \texttt{Multilingual-E5-large}\footnote{\url{https://huggingface.co/intfloat/multilingual-e5-large}} model, selected for its robust performance in semantic retrieval tasks. For Stage III arbitration and baseline comparisons, we integrate a spectrum of open-source LLMs including \texttt{Llama-3-8B}\footnote{\url{https://huggingface.co/meta-llama/Meta-Llama-3-8B}}, \texttt{Mistral-7B}\footnote{\url{https://huggingface.co/mistralai/Mistral-7B-v0.1}}, \texttt{Gemma-7B}\footnote{\url{https://huggingface.co/google/gemma-7b}}, \texttt{Qwen2.5-0.5B}\footnote{\url{https://huggingface.co/Qwen/Qwen2.5-0.5B}}, \texttt{TinyLlama-1.1B}\footnote{\url{https://huggingface.co/TinyLlama/TinyLlama-1.1B-Chat-v1.0}}, and \texttt{Llama-2-13B}\footnote{\url{https://huggingface.co/meta-llama/Llama-2-13b-chat}}. Additionally, we evaluate performance against state-of-the-art proprietary APIs, specifically \texttt{GPT-4o-mini} and \texttt{DeepSeek-chat}\footnote{\url{https://huggingface.co/deepseek-ai}}, to benchmark our framework against industry standards.

\paragraph{Competing Baselines}
We benchmark \textsc{MCP-Guard} against three open-source defenses: 
\textit{SafeMCP} \cite{fang2025identify} layers regex whitelisting with LLaMA-Guard and OpenAI Moderation to block poisoning and command injections; 
\textit{MCP-Shield} \cite{MCPShield2024} combines rule-based static analysis with optional Claude-powered semantic checks to detect shadowing and data exfiltration; 
and \textit{MCP-Scan} \cite{mcp-scan} integrates offline audits with live proxy monitoring for real-time threat detection. 
All baselines utilize \texttt{GPT-4o-mini} as the unified backend, with \emph{suspicious} and \emph{malicious} outputs merged into a single \emph{unsafe} class for consistent binary evaluation.

\begin{table*}[t]
\centering
\scriptsize
\caption{Main experimental results. (a) Comparative analysis of \textsc{MCP-Guard} against state-of-the-art baselines and internal ablation on \textsc{MCP-AttackBench}. (b) Generalizability and efficiency assessment across external benchmarks (\textit{AgentDefense}, \textit{MCPSecBench}, and \textit{RAS-Eval}).}
\label{tab:main_results}

\begin{subtable}[t]{0.48\linewidth}
\centering
\caption{Performance on \textsc{MCP-AttackBench}.}
\label{tab:sub_component}
\sisetup{
    tight-spacing = true,
    round-mode = places,
    round-precision = 1,
    detect-weight = true
}
\setlength{\tabcolsep}{2pt}
\renewcommand{\arraystretch}{1.15}
\begin{tabular}{l *{4}{S[table-format=2.1]} S[table-format=4.1]}
\toprule
\textbf{Method} & {\textbf{Acc}} & {\textbf{Prec}} & {\textbf{Rec}} & {\textbf{F1}} & {\textbf{Time}} \\
& {(\%)} & {(\%)} & {(\%)} & {(\%)} & {(ms)} \\
\midrule
\rowcolor{gray!10} \multicolumn{6}{l}{\textbf{\textsc{MCP-Guard} Internal Stages}} \\
\quad Pattern (Stage I)          & 74.6 & \bfseries 97.7 & 38.9  & 55.6 & \bfseries 1.8   \\
\quad Learnable (Stage II)        & \bfseries 96.0 & 96.7 & 93.5  & \bfseries 95.1 & 55.1  \\
\quad \texttt{GPT-4o-mini} (Stage III) & 95.4 & 92.1 & \bfseries 97.0 & 94.5 & 788.4 \\
\quad \texttt{TinyLlama:1.1B} (Stage III) & 60.8 & 59.6 & 13.7  & 22.2 & 628.8 \\
\midrule
\rowcolor{gray!10} \multicolumn{6}{l}{\textbf{Competing Baselines}} \\
\quad MCP-Scan (\texttt{GPT-4o-mini})   & 94.0 & \bfseries 99.7 & 85.7  & 92.2 & 613.2 \\
\quad SafeMCP (\texttt{GPT-4o-mini})    & 79.3 & 66.9 & 98.1  & 79.6 & 2292.8 \\
\quad MCP-Shield (\texttt{GPT-4o-mini}) & 53.5 & 46.7 & 93.5  & 62.2 & 6212.3 \\
\rowcolor{cyan!10}
\quad \textsc{MCP-Guard} (\texttt{GPT-4o-mini}) & \bfseries 96.0 & 91.5 & \bfseries 99.5 & \bfseries 95.4 & \bfseries 505.9 \\
\bottomrule
\end{tabular}
\end{subtable}
\hfill
\begin{subtable}[t]{0.48\linewidth}
\centering
\caption{Performance on external defense benchmarks.}
\label{tab:external_benchmarks}
\sisetup{
    tight-spacing = true,
    round-mode = places,
    round-precision = 2,
    detect-weight = true
}
\setlength{\tabcolsep}{2pt}
\renewcommand{\arraystretch}{1.15}
\begin{tabular}{l *{4}{S[table-format=3.2]} S[table-format=4.2]}
\toprule
\textbf{Backbone / Benchmark} & {\textbf{Acc}} & {\textbf{Prec}} & {\textbf{Rec}} & {\textbf{F1}} & {\textbf{Time}} \\
& {(\%)} & {(\%)} & {(\%)} & {(\%)} & {(ms)} \\
\midrule
\rowcolor{gray!10} \multicolumn{6}{l}{\textbf{\textsc{MCP-Guard} (\texttt{Llama3-8B})}} \\
\quad AgentDefense & 93.10 & 100.00 & 93.10 & 96.43 & 55.98  \\
\quad MCPSecBench  & 90.00 & 100.00 & 90.00 & 94.74 & 47.57  \\
\quad RAS-Eval     & 96.84 & 99.30  & 97.46 & 98.37 & 152.62 \\
\rowcolor{cyan!10} \quad \textbf{Average} & 93.31 & 99.77 & 93.52 & 96.51 & 85.39  \\
\midrule
\rowcolor{gray!10} \multicolumn{6}{l}{\textbf{\textsc{MCP-Guard} (\texttt{Deepseek-chat})}} \\
\quad AgentDefense & 96.87 & 100.00 & 96.87 & 98.51 & 192.87 \\
\quad MCPSecBench  & 90.00 & 100.00 & 90.00 & 94.74 & 166.23 \\
\quad RAS-Eval     & 96.84 & 99.30  & 97.46 & 98.37 & 403.22 \\
\rowcolor{cyan!10} \quad \textbf{Average} & 94.57 & 99.77 & 94.78 & 97.21 & 254.11 \\
\bottomrule
\end{tabular}
\end{subtable}
\end{table*}

\begin{table}[t]
\centering
\caption{Comprehensive Performance and Efficiency Gain: Standalone LLMs vs. MCP-Guard Framework}
\label{tab:performance_gain_complete}
\scriptsize
\setlength{\tabcolsep}{1pt}  
\renewcommand{\arraystretch}{1.15}
\begin{tabular}{l
  | rr
  | rr
  | r r}
\toprule
\textbf{Base Model}
  & \multicolumn{2}{c|}{\textbf{Standalone}}
  & \multicolumn{2}{c|}{\textbf{MCP-Guard}}
  & \multicolumn{2}{c}{\textbf{Net Improvement}} \\
  & F1 (\%) & Time (ms) & F1 (\%) & Time (ms) & $\Delta$ F1 & Speedup \\
\midrule
\texttt{GPT-4o-mini}     & \textbf{94.5} & 788.4  & \textbf{95.4} & 505.9  & +0.9  & 1.56$\times$ \\
\texttt{Deepseek-chat}   & 90.8          & 3358.0 & 93.1          & 1988.2 & +2.3  & 1.69$\times$ \\
\texttt{Mistral:7B}       & 76.3          & 435.4  & 89.6          & 157.3  & +13.3 & 2.77$\times$ \\
\texttt{Qwen2.5:0.5B}     & 76.9          & \textbf{157.9} & 92.7  & 143.7  & +15.8 & 1.10$\times$ \\
\texttt{Llama3:8B}        & 57.8          & 167.6  & \textbf{95.4} & \textbf{91.5} & +37.6 & 1.83$\times$ \\
\texttt{TinyLlama:1.1B}   & 22.2          & 628.8  & 83.4          & 333.2  & \textbf{+61.2} & 1.89$\times$ \\
\texttt{Llama2:13B}       & 51.7          & 1490.2 & 76.2          & 232.5  & +24.5 & \textbf{6.41}$\times$ \\
\texttt{Gemma:7B}         & 55.3          & 413.7  & 86.7          & 194.7  & +31.4 & 2.12$\times$ \\
\midrule
\rowcolor{cyan!10}
\textbf{Average}         & 65.7          & 930.0  & 89.1          & 455.9  & {+23.4} & {2.04}$\times$ \\
\bottomrule
\end{tabular}
\end{table}

\subsection{Experimental Results and Analysis}
\label{sec:results}

\subsubsection{RQ1: Effectiveness}
To answer \textbf{RQ1}, we evaluate the detection capability of \textsc{MCP-Guard} against state-of-the-art baselines across diverse threat landscapes. As illustrated by the performance trajectory in Figure~\ref{fig:comp_sys_evo} and the comparative metrics in Table~\ref{tab:main_results}, our framework effectively establishes an {optimal Pareto frontier}. 

\paragraph{Competitive General Performance}
Table~\ref{tab:sub_component} reports the comprehensive detection performance of \textsc{MCP-Guard} compared to existing baselines on the \textsc{MCP-AttackBench} dataset, \textsc{MCP-Guard} achieves the \textbf{optimal Pareto frontier} with a peak F1-score of 95.4\%, significantly outperforming baselines while maintaining lower latency than heavy-model counterparts. It balances high precision (91.5\%) and superior recall (99.5\%), avoiding \textit{MCP-Scan}'s low recall (85.7\%) that misses stealthy attacks and \textit{SafeMCP}'s low precision (66.9\%) that causes excessive false alarms.

\paragraph{Generalization on External Benchmarks}
To assess robustness beyond \textsc{MCP-AttackBench}, we evaluated backend models on \textit{AgentDefense}, \textit{MCPSecBench}, and \textit{RAS-Eval} (Table~\ref{tab:external_benchmarks}). Using \texttt{Deepseek-chat} as Stage III yields an average F1-score of 97.21\%, peaking at 98.51\% on \textit{AgentDefense}. The lighter \texttt{Llama-3-8B} achieves 96.51\% average F1 with markedly lower latency (85.39ms), further validating that our architecture ensures high-security standards across \textbf{varying model scales and benchmarks}.

\subsubsection{RQ2: Architecture \& Efficiency }

To address \textbf{RQ2}, we evaluate whether the cascaded design of \textsc{MCP-Guard} successfully reconciles the conflict between rigorous security inspection and the low-latency requirements of real-time agentic workflows.

\paragraph{Stage I's Fail-Fast Mechanism} Table~\ref{tab:sub_component} shows that \textit{Pattern (Stage I)} serves as a high-confidence sieve: it achieves 97.7\% precision but only 38.9\% recall, confirming its effectiveness in rapidly filtering explicit syntactic attacks \textbf{efficiently in 1.8 ms on average}, while revealing its limitations against semantic threats.

\paragraph{Stage II's Semantic Neural Detection} 
Addressing the limited recall of Stage I (38.9\%), Stage II leverages a fine-tuned E5 embedding model to capture obfuscated semantic threats. Full-parameter fine-tuning on \textsc{MCP-AttackBench} overcomes the domain misalignment of standard embeddings (65.37\% accuracy), propelling the F1-score from \textbf{55.6\% (Stage I) to 95.1\% (Stage II)} with 96.01\% accuracy (Table~\ref{tab:sub_component}). This substantial gain confirms the neural component's critical role in identifying complex attacks that evade rigid syntactic filters.

\paragraph{Speedup Against LLMs Standalone (Stage III)} Table \ref{tab:sub_component} shows that the \textsc{MCP-Guard}(\texttt{GPT-4o-mini})  operates with an average latency of {505.9 ms}. This represents a {1.56$\times$ speedup} compared to a standalone \texttt{GPT-4o-mini} (788.4 ms) and a massive {12$\times$ speedup} compared to \textit{MCP-Shield} (6212 ms).  As detailed further in Table~\ref{tab:performance_gain_complete} and Figure \ref{fig:performance_grid}, the framework reduces inference latency by half, maintaining a \textbf{2.04$\times$ average speedup} against LLMs Standalone (Stage III) .

\paragraph{$\Delta$ F1 Against LLMs Standalone (Stage III)}   As shown in Figure~\ref{fig:comp_sys_evo}, Figure~\ref{fig:performance_grid}, and Table~\ref{tab:performance_gain_complete}, \textsc{MCP-Guard} delivers a consistent ``lifting effect'' across diverse backbones, effectively patching weaker models. It boosts \texttt{TinyLlama-1.1B}'s F1-score by 61.2\%, achieving an \textbf{Avg. $\Delta$ F1=23.4} against LLMs Standalone (Stage III).

\section{Conclusion} 

The standardization of the MCP empowers LLM agents but exposes them to critical vulnerabilities. To address this, we introduced \textsc{MCP-Guard}, a multi-stage defense framework that reconciles high-precision security with real-time latency through a cascaded architecture of Lightweight Syntactic Filtering (Stage I), Semantic Neural Detection with E5 text
embedding (Stage II), and Cognitive Arbitration (Stage III). Our evaluation demonstrates that \textsc{MCP-Guard} effectively breaks the efficiency-robustness trade-off, achieving an optimal F1-score of 95.4\% and a $2.04\times$ speedup over monolithic defenses. Extensive validation on external benchmarks, such as \textit{AgentDefense} and \textit{RAS-Eval}, further confirms the framework's generalization capabilities across diverse threat landscapes. As MCP evolves into a universal connectivity layer, \textsc{MCP-Guard} establishes a foundational, scalable blueprint for securing the agentic AI supply chain.
\section*{Limitations}
Despite the robust performance of \textsc{MCP-Guard}, several limitations remain inherent to its current design and evaluation scope:

\paragraph{Protocol Dependency and Evolution} 
Our framework is tightly coupled with the current specification of the Model Context Protocol. While Stage I's regex patterns are hot-updateable, fundamental changes to the MCP transport layer (e.g., a shift from JSON-RPC to a binary protocol) would necessitate significant re-engineering of the parsing logic. Additionally, our evaluation primarily focuses on text-based payloads. As MCP evolves to support multi-modal data transfer (e.g., image or audio buffers), our text-centric embedding models (Stage II) may require retraining to detect adversarial perturbations in non-textual modalities.

\paragraph{Latency vs. Security Trade-off} 
Although \textsc{MCP-Guard} achieves a $2.04\times$ speedup over monolithic defenses, the average latency of 505.9 ms may still be prohibitive for ultra-low-latency applications, such as high-frequency trading agents or real-time industrial control systems.

\section*{Ethical Considerations}
\paragraph{Dual-Use Risks of MCP-AttackBench} 
 We acknowledge the risk that this dataset could be misused to train more sophisticated attack agents. To mitigate this, we will release the dataset under a restrictive research-only license and have sanitized the samples to remove personally identifiable information (PII) and live credentials, ensuring they serve as educational artifacts rather than ready-to-use exploit kits.

\paragraph{Privacy and Data Inspection} 
\textsc{MCP-Guard} operates as a middleware proxy that inspects the semantic content of tool invocations. This necessitates the decryption and analysis of potentially sensitive user data (e.g., file contents, database queries). In enterprise deployments, this centralized inspection point introduces a new privacy target. We emphasize that \textsc{MCP-Guard} should be deployed within the user's trusted infrastructure (e.g., local VPC or on-premise), and we recommend configuring data retention policies that discard payload content immediately after inference to prevent the accumulation of sensitive logs.

\bibliography{acl_CRA}

@inproceedings{jing2025mcip,
  title={Mcip: Protecting mcp safety via model contextual integrity protocol},
  author={Jing, Huihao and Li, Haoran and Hu, Wenbin and Hu, Qi and Heli, Xu and Chu, Tianshu and Hu, Peizhao and Song, Yangqiu},
  booktitle={Proceedings of the 2025 Conference on Empirical Methods in Natural Language Processing},
  pages={1177--1194},
  year={2025}
}

@article{radosevich2025mcp,
  title={Mcp safety audit: Llms with the model context protocol allow major security exploits},
  author={Radosevich, Brandon and Halloran, John},
  journal={arXiv preprint arXiv:2504.03767},
  year={2025}
}

@article{yang2025mcpsecbench,
  title={Mcpsecbench: A systematic security benchmark and playground for testing model context protocols},
  author={Yang, Yixuan and Wu, Daoyuan and Chen, Yufan},
  journal={arXiv preprint arXiv:2508.13220},
  year={2025}
}

@article{li2024gentel,
  title={Gentel-safe: A unified benchmark and shielding framework for defending against prompt injection attacks},
  author={Li, Rongchang and Chen, Minjie and Hu, Chang and Chen, Han and Xing, Wenpeng and Han, Meng},
  journal={arXiv preprint arXiv:2409.19521},
  year={2024}
}

@article{wang2025mpma,
title={MPMA: Preference Manipulation Attack Against Model Context Protocol},
author={Wang, Zihan and Li, Hongwei and Zhang, Rui and Liu, Yu and Jiang, Wenbo and Fan, Wenshu and Zhao, Qingchuan and Xu, Guowen},
journal={arXiv preprint arXiv:2506.02040},
year={2025}
}

@article{narajala2025securing,
  title={Securing GenAI Multi-Agent Systems Against Tool Squatting: A Zero Trust Registry-Based Approach},
  author={Narajala, Vineeth Sai and Huang, Ken and Habler, Idan},
  journal={arXiv preprint arXiv:2504.19951},
  year={2025},
  url={https://arxiv.org/abs/2504.19951}
}

@misc{mcp-scan,
  author       = {Invariant-Labs},
  title        = {MCP-Scan: A Lightweight Security Detection Framework},
  year         = {2024},
  howpublished = {\url{https://github.com/invariantlabs-ai/mcp-scan}},
  note         = {Accessed: 2025-07-31}
}

@article{kumar2025mcp,
  title={Mcp guardian: A security-first layer for safeguarding mcp-based ai system},
  author={Kumar, Sonu and Girdhar, Anubhav and Patil, Ritesh and Tripathi, Divyansh},
  journal={arXiv preprint arXiv:2504.12757},
  year={2025}
}

@article{wang2025mcpguard,
  title={Mcpguard: Automatically detecting vulnerabilities in mcp servers},
  author={Wang, Bin and Liu, Zexin and Yu, Hao and Yang, Ao and Huang, Yenan and Guo, Jing and Cheng, Huangsheng and Li, Hui and Wu, Huiyu},
  journal={arXiv preprint arXiv:2510.23673},
  year={2025}
}

@article{bhatt2025etdi,
  title={ETDI: Mitigating Tool Squatting and Rug Pull Attacks in Model Context Protocol (MCP) by using OAuth-Enhanced Tool Definitions and Policy-Based Access Control},
  author={Bhatt, Manish and Narajala, Vineeth Sai and Habler, Idan},
  journal={arXiv preprint arXiv:2506.01333},
  year={2025},
  url={https://arxiv.org/abs/2506.01333}
}

@article{invariant2025poisoning,
  title={Mcp security notification: Tool poisoning attacks},
  author={Beurer-Kellner, Luca and Fischer, Marc},
  journal={Invariant Labs Blog},
  year={2025}
}

@misc{cloudflare2025mcp,
  title={MCP Connectors on Cloudflare Workers},
  author={Cloudflare},
  year={2025},
  howpublished={Cloudflare Blog},
  url={https://blog.cloudflare.com/building-ai-agents-with-mcp-authn-authz-and-durable-objects}
}

@misc{sanna2025agentdefensebench,
  title={AgentDefense-Bench: A Security Benchmark for MCP-Based AI Agents},
  author={Sanna, Arun},
  year={2025},
  url={https://github.com/arunsanna/AgentDefense-Bench}
}

@article{guo2025systematic,
  title={Systematic analysis of mcp security},
  author={Guo, Yongjian and Liu, Puzhuo and Ma, Wanlun and Deng, Zehang and Zhu, Xiaogang and Di, Peng and Xiao, Xi and Wen, Sheng},
  journal={arXiv preprint arXiv:2508.12538},
  year={2025}
}

@article{wang2022text,
  title={Text embeddings by weakly-supervised contrastive pre-training},
  author={Wang, Liang and Yang, Nan and Huang, Xiaolong and Jiao, Binxing and Yang, Linjun and Jiang, Daxin and Majumder, Rangan and Wei, Furu},
  journal={arXiv preprint arXiv:2212.03533},
  year={2022}
}

@article{brett2025gateway,
title={Simplified and Secure MCP Gateways for Enterprise AI Integration},
author={Brett, Ivo},
journal={Preprint},
year={2025},
note={Available at \url{https://independent.academia.edu/ivobrett}}
}

@article{hou2025landscape,
  title={Model Context Protocol (MCP): Landscape, Security Threats, and Future Research Directions},
  author={Hou, Xinyi and Zhao, Yanjie and Wang, Shenao and Wang, Haoyu},
  journal={arXiv preprint arXiv:2503.23278},
  year={2025},
  note={Huazhong University of Science and Technology, China}
}

@article{fang2025identify,
  title={We Should Identify and Mitigate Third-Party Safety Risks in MCP-Powered Agent Systems},
  author={Fang, Junfeng and Yao, Zijun and Wang, Ruipeng and Ma, Haokai and Wang, Xiang and Chua, Tat-Seng},
  journal={arXiv preprint arXiv:2506.13666v1},
  year={2025},
  url={https://arxiv.org/abs/2506.13666v1}
}

@article{zhang2025meraser,
  title={MEraser: An Effective Fingerprint Erasure Approach for Large Language Models},
  author={Zhang, Jingxuan and Xu, Zhenhua and Hu, Rui and Xing, Wenpeng and Zhang, Xuhong and Han, Meng},
  journal={arXiv preprint arXiv:2506.12551},
  year={2025}
}

@article{xu2025rap,
  title={RAP-SM: Robust Adversarial Prompt via Shadow Models for Copyright Verification of Large Language Models},
  author={Xu, Zhenhua and Wang, Zhebo and Li, Maike and Xing, Wenpeng and Hu, Chunqiang and Zhi, Chen and Han, Meng},
  journal={arXiv preprint arXiv:2505.06304},
  year={2025}
}

@article{xing2025latent,
  title={Latent Fusion Jailbreak: Blending Harmful and Harmless Representations to Elicit Unsafe LLM Outputs},
  author={Xing, Wenpeng and Li, Mohan and Hu, Chunqiang and Zhang, Haitao XuNingyu and Lin, Bo and Han, Meng},
  journal={arXiv preprint arXiv:2508.10029},
  year={2025}
}

@article{xu1906insty,
  title={Insty: a robust multi-level crossgranularity fingerprint embedding algorithm for multi-turn dialogue in large language models},
  author={Xu, Zhenhua and Han, Meng and Yue, Xubin and Xing, Wenpeng},
  journal={SCIENTIA SINICA Informationis},
  volume={55},
  number={8},
  year={1906}
}

@article{xu2025copyright,
  title={Copyright Protection for Large Language Models: A Survey of Methods, Challenges, and Trends},
  author={Xu, Zhenhua and Yue, Xubin and Wang, Zhebo and Liu, Qichen and Zhao, Xixiang and Zhang, Jingxuan and Zeng, Wenjun and Xing, Wengpeng and Kong, Dezhang and Lin, Changting and others},
  journal={arXiv preprint arXiv:2508.11548},
  year={2025}
}

@article{yue2025pree,
  title={Pree: Towards harmless and adaptive fingerprint editing in large language models via knowledge prefix enhancement},
  author={Yue, Xubin and Xu, Zhenhua and Xing, Wenpeng and Yu, Jiahui and Li, Mohan and Han, Meng},
  journal={Preprint},
  year={2025}
}

@inproceedings{xu2025evertracer,
  title={Evertracer: Hunting stolen large language models via stealthy and robust probabilistic fingerprint},
  author={Xu, Zhenhua and Han, Meng and Xing, Wenpeng},
  booktitle={Proceedings of the 2025 Conference on Empirical Methods in Natural Language Processing},
  pages={7019--7042},
  year={2025}
}

@inproceedings{li2025optimizing,
  title={Optimizing and Attacking Embodied Intelligence: Instruction Decomposition and Adversarial Robustness},
  author={Li, Minghao and Xing, Wenpeng and Liu, Yong and Zhang, Wei and Han, Meng},
  booktitle={2025 IEEE International Conference on Multimedia and Expo (ICME)},
  pages={1--6},
  year={2025},
  organization={IEEE}
}

@article{xing2025towards,
  title={Towards robust and secure embodied ai: A survey on vulnerabilities and attacks},
  author={Xing, Wenpeng and Li, Minghao and Li, Mohan and Han, Meng},
  journal={arXiv preprint arXiv:2502.13175},
  year={2025}
}

@misc{MCPShield2024,
  author       = {Nikita Kryzhanouski},
  title        = {MCP-Shield: Safety-Constrained Multi-Agent Path Planning},
  year         = {2024},
  howpublished = {\url{https://github.com/riseandignite/mcp-shield}},
  note         = {Accessed: 2025-07-31}
}

@misc{MCP,
  author       = {Anthropic},
  title        = {Introducing the Model Context Protocol},
  year         = {2025},
  howpublished = {\url{https://www.anthropic.com/news/model-context-protocol}},
  note         = {Accessed: 2025-08-1}
}

@article{fu2025ras,
  title={RAS-Eval: A Comprehensive Benchmark for Security Evaluation of LLM Agents in Real-World Environments},
  author={Fu, Yuchuan and Yuan, Xiaohan and Wang, Dongxia},
  journal={arXiv preprint arXiv:2506.15253},
  year={2025}
}

\appendix

\section{Complete Decision Path of MCP-GUARD}
\label{sec:setup_apdx}

Figure~\ref{fig:decision_path} provides a detailed view of the complete decision workflow of \textsc{MCP-Guard}. Stage I performs lightweight static scanning with a fail-fast block for overt threats. Requests passing Stage I proceed to Stage II, where the fine-tuned E5 model computes a malice probability score $P(y|x)$. Only ambiguous predictions (e.g., 0.45 < $P(y|x)$ < 0.55) trigger Stage III LLM arbitration, which outputs Safe (S), Unsafe (U), or Uncertain (U$_c$). Both non-ambiguous cases from Stage II and uncertain verdicts from Stage III fallback to the efficient neural threshold $T_u$ for final decision, reserving expensive LLM reasoning for the most challenging inputs while achieving sub-millisecond average overhead for the majority of traffic.

\begin{figure*}[t]
\centering
\small
\begin{tikzpicture}[
    node distance=0.9cm and 1.8cm,
    font=\scriptsize\sffamily,
    >=Stealth,
    base/.style={draw=black, thick, rounded corners=2pt, align=center, minimum width=2.8cm, minimum height=0.7cm, fill=white},  
    decision/.style={diamond, draw=black, thick, fill=white, aspect=2.8, inner sep=1pt, align=center},  
    terminal/.style={base, fill=gray!12, dashed, thick, rounded corners=4pt, minimum width=2.6cm, draw=black, thick},  
    block/.style={base, fill=red!15, draw=black, thick},           
    allow/.style={base, fill=green!15, draw=black, thick},         
    stage1/.style={base, fill=green!8, draw=black, thick},         
    stage2/.style={base, fill=blue!12, draw=black, thick},         
    stage3/.style={base, fill=orange!15, draw=black, thick},       
    fallback/.style={decision, fill=green!5, draw=black, thick},   
    route/.style={->, thick, draw=gray!55},
    danger/.style={->, thick, draw=red!65},    
    safe/.style={->, thick, draw=green!65},    
    dashroute/.style={->, dashed, thick, draw=gray!55}  
]
    \node (start) [terminal] {User Prompt / Tool Call};
    \node (s1) [stage1, below=of start] {\textbf{Stage I}: Static Scanning\\(Regex \& Pattern)};
    \node (dec1) [decision, below=of s1] {Hit Rule?};
    \node (block_s1) [block, left=1.8cm of dec1] {\textbf{BLOCK}\\(Fail-Fast)};
    \node (s2) [stage2, below=of dec1] {\textbf{Stage II}: Neural Detection\\(E5 Model)};
    \node (dec_amb) [decision, below=of s2] {Ambiguous?\\(e.g., 0.45 < $P$ < 0.55)};
   
    \node (dec_fallback) [fallback, right=2.2cm of dec_amb] {Neural Fallback:\\$P(y|x) > T_u$?};
    \node (s3) [stage3, below=of dec_amb] {\textbf{Stage III}: LLM Arbitration\\(Zero-shot)};
    \node (dec3) [decision, below=of s3] {LLM Verdict?};
    \node (block_final) [block, below=1.2cm of dec3, xshift=-1cm] {\textbf{BLOCK}};
    \node (allow_final) [allow, right=2cm of block_final] {\textbf{ALLOW}\\(To Server)};

    \draw [route] (start) -- (s1);
    \draw [route] (s1) -- (dec1);
    \draw [danger] (dec1) -- node[above, font=\scriptsize] {Yes} (block_s1);
    \draw [route] (dec1) -- node[right, font=\scriptsize] {No} (s2);
    \draw [route] (s2) -- (dec_amb);
    \draw [route] (dec_amb) -- node[left, font=\scriptsize] {Yes} (s3);
    \draw [dashroute] (dec_amb) -- node[above, font=\scriptsize] {No} (dec_fallback);
    \draw [route] (s3) -- (dec3);
    \draw [danger] (dec3) -- node[left, font=\scriptsize, near start] {Unsafe (U)} (block_final);
    \draw [safe] (dec3) -- node[above left, font=\scriptsize, pos=0.6] {Safe (S)} (allow_final);
    \draw [dashroute] (dec3) -| node[above, font=\scriptsize, pos=0.3] {Uncertain (U$_c$)} (dec_fallback);
    \draw [danger] (dec_fallback) -- (block_final);
    \draw [safe] (dec_fallback) -- node[above, font=\scriptsize] {No} (allow_final);
    \draw [dashroute, ->, out=20, in=90, looseness=1.3] (s2.east) to node[above, sloped, font=\tiny, text=gray!80] {Score $P(y|x)$} (dec_fallback.north);

    \begin{scope}[on background layer]
        \node [fill=gray!6, rounded corners, inner sep=8pt, fit=(s1) (dec1) (block_s1)]
              [label={[anchor=north east, font=\tiny\itshape, color=gray!70]north east:Lightweight}] {};
        \node [fill=gray!4, rounded corners, inner sep=10pt, fit=(s2) (dec_amb) (s3) (dec3) (dec_fallback)]
              [label={[anchor=south east, font=\tiny\itshape, color=gray!70]south east:Heavyweight (Conditional)}] {};
    \end{scope}
\end{tikzpicture}
\caption{{The Decision Path of \textsc{MCP-Guard}.} The workflow explicitly shows that Stage III LLM arbitration is triggered only for ambiguous cases from Stage II. Non-ambiguous cases and LLM uncertainty both fallback to the efficient neural score ($P(y|x) > T_u$), ensuring low average latency while maintaining high accuracy.}
\label{fig:decision_path}
\end{figure*}

\section{Stage I: Lightweight Static Scanning by Pattern-based Detectors}
\label{sec:stage1}

\begin{figure*}[t!]
    \centering
    \begin{subfigure}[b]{0.24\textwidth}
        \centering
        \includegraphics[width=\textwidth]{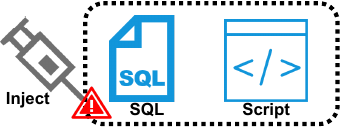}
        \caption{SQL Injection}
        \label{fig:attack_sql}
    \end{subfigure}
    \hfill
    \begin{subfigure}[b]{0.24\textwidth}
        \centering
        \includegraphics[width=\textwidth]{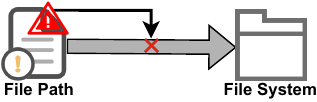}
        \caption{Sensitive Files}
        \label{fig:attack_file}
    \end{subfigure}
    \hfill
    \begin{subfigure}[b]{0.24\textwidth}
        \centering
        \includegraphics[width=\textwidth]{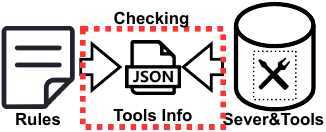}
        \caption{Shadow Hijack}
        \label{fig:attack_shadow}
    \end{subfigure}
    \hfill
    \begin{subfigure}[b]{0.24\textwidth}
        \centering
        \includegraphics[width=0.8\textwidth]{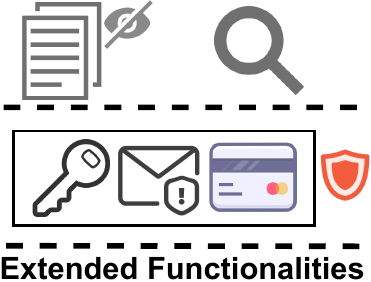}
        \caption{Prompt Injection}
        \label{fig:attack_prompt}
    \end{subfigure}
    \vspace{8pt} 
    \begin{subfigure}[b]{0.24\textwidth}
        \centering
        \includegraphics[width=\textwidth]{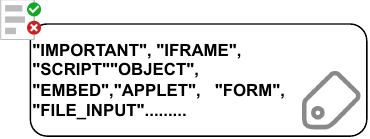}
        \caption{Important Tag}
        \label{fig:attack_tag}
    \end{subfigure}
    \hfill
    \begin{subfigure}[b]{0.24\textwidth}
        \centering
        \includegraphics[width=\textwidth]{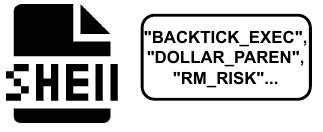}
        \caption{Shell Injection}
        \label{fig:attack_shell}
    \end{subfigure}
    \hfill
    \begin{subfigure}[b]{0.24\textwidth}
        \centering
        \includegraphics[width=\textwidth]{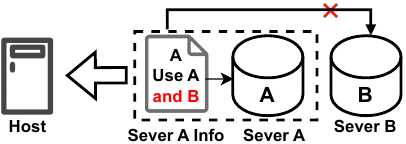}
        \caption{Cross-Origin}
        \label{fig:attack_origin}
    \end{subfigure}
    \caption{\textbf{Taxonomy of Attack Vectors in Stage 1.} The figure illustrates the diverse set of malicious patterns captured by our static scanning mechanism, ranging from traditional injection attacks (a, f) to LLM-specific vulnerabilities like Prompt Injection (d) and Shadow Hijacking (c).}
    \label{fig:stage1_detectors_grid}
\end{figure*}

This stage employs a suite of high-performance, pattern-based detectors designed to intercept obvious security threats at the earliest possible phase. By filtering common attack vectors before they reach computationally expensive neural models, the pipeline significantly minimizes total inference latency. If any high-confidence rule is triggered, the system executes a "fail-fast" block, optimizing resource allocation. The visual patterns and execution flows for these detectors are systematically illustrated in the grid in Figure~\ref{fig:stage1_detectors_grid}.

\begin{enumerate}
    \item \textbf{SQL Injection Detector}: As depicted in Figure~\ref{fig:attack_sql}, this module monitors for traditional injection vectors by matching patterns associated with SQL administrative commands and script-based triggers:
    \begin{center}
    \begin{minipage}{0.85\columnwidth}
    \begin{lstlisting}[style=regexstyle, frame=single, basicstyle=\ttfamily\scriptsize]
(--|\b{OR}\b|\b{AND}\b).*(=|LIKE), <\s*script\b
    \end{lstlisting}
    \end{minipage}
    \end{center}
    
    \item \textbf{Sensitive File Detector}: This detector (see Figure~\ref{fig:attack_file}) acts as a data loss prevention (DLP) mechanism, intercepting unauthorized attempts to access system-level directories or environment configurations:
    \begin{center}
    \begin{minipage}{0.85\columnwidth}
    \begin{lstlisting}[style=regexstyle, frame=single, basicstyle=\ttfamily\scriptsize]
\.ssh/, \.env\b, /etc/passwd
    \end{lstlisting}
    \end{minipage}
    \end{center}
    
    \item \textbf{Shadow Hijack Detector}: To mitigate the masquerading risks shown in Figure~\ref{fig:attack_shadow}, this detector identifies spoofed server responses or hidden tool invocation instructions that bypass standard intent parsing:
    \begin{center}
    \begin{minipage}{0.85\columnwidth}
    \begin{lstlisting}[style=regexstyle, frame=single, basicstyle=\ttfamily\scriptsize]
\bspoofed\s+call\b, \bfake\s+server\b
    \end{lstlisting}
    \end{minipage}
    \end{center}
    
    \item \textbf{Prompt Injection Detector}: This multi-stage filter handles complex adversarial prompts illustrated in Figure~\ref{fig:attack_prompt}. It combines case-insensitive keyword filtering with dynamic RegEx for obfuscated command identification:
    \begin{center}
    \begin{minipage}{0.85\columnwidth}
    \begin{lstlisting}[style=regexstyle, frame=single, basicstyle=\ttfamily\scriptsize]
\bignore\s+previous\b, \bexecute\s+hidden\b
    \end{lstlisting}
    \end{minipage}
    \end{center}
    
    \item \textbf{Important Tag Detector}: Specifically designed to expose the hidden carriers within tool descriptions (Figure~\ref{fig:attack_tag}), this module captures the \texttt{<IMPORTANT>} tag and related HTML-based injection tags:
    \begin{center}
    \begin{minipage}{0.85\columnwidth}
    \begin{lstlisting}[style=regexstyle, frame=single, basicstyle=\ttfamily\scriptsize]
<\s*important\b, <\s*iframe\b, <\s*form\b
    \end{lstlisting}
    \end{minipage}
    \end{center}
    
    \item \textbf{Shell Injection Detector}: Leveraging the patterns shown in Figure~\ref{fig:attack_shell}, this detector utilizes heuristic and lexical analysis to identify high-risk shell command sequences in user-provided inputs:
    \begin{center}
    \begin{minipage}{0.85\columnwidth}
    \begin{lstlisting}[style=regexstyle, frame=single, basicstyle=\ttfamily\scriptsize]
\b(sh|bash|curl|rm|wget|chmod)\b
    \end{lstlisting}
    \end{minipage}
    \end{center}
    
    \item \textbf{Cross-Origin Detector}: Guided by the logic in Figure~\ref{fig:attack_origin}, this detector validates external server references against a dynamic whitelist to prevent unauthorized cross-origin data exfiltration:
    \begin{center}
    \begin{minipage}{0.85\columnwidth}
    \begin{lstlisting}[style=regexstyle, frame=single, basicstyle=\ttfamily\scriptsize]
\bexternal-server\b, \bthird-party-api\b
    \end{lstlisting}
    \end{minipage}
    \end{center}
\end{enumerate}

\section{End-to-End Efficiency and Performance Gains}
Table~\ref{tab:performance_gain_complete2} presents a comprehensive comparison between standalone LLM arbitration (Stage III only) and the full \textsc{MCP-Guard} pipeline across eight representative base models.
The full framework achieves an average F1-score of 89.1\% (+23.4\% absolute improvement) and an average latency of 455.9 ms—a 2.04$\times$ speedup over standalone LLM defenses (average 930.0 ms). Gains are particularly pronounced for smaller and older models: \texttt{TinyLlama-1.1B} improves by +61.2 F1 points with 1.89$\times$ speedup, while \texttt{Llama2-13B} yields the highest speedup (6.41$\times$) alongside +24.5 F1 points. Even high-performing models like \texttt{GPT-4o-mini} benefit from reduced latency (1.56$\times$) and slight accuracy gains (+0.9 F1).
Across all models, recall increases substantially (from 70.2\% to 98.5\% on average), reflecting the pipeline's ability to preserve sensitive threat detection while the cascaded design dramatically lowers computational overhead. Compared to prior work such as \textit{MCP-Shield} (reported 6212 ms latency), \textsc{MCP-Guard} delivers up to 13.6$\times$ overall speedup, demonstrating the practical value of layered, efficiency-aware defense.

\begin{table*}[t]
\centering
\caption{Comprehensive Performance and Efficiency Gain: Standalone LLMs vs. MCP-Guard Framework}
\label{tab:performance_gain_complete2}
\setlength{\tabcolsep}{3pt} 
\renewcommand{\arraystretch}{1.1}
\begin{tabular}{l|ccccc|ccccc|cc}
\toprule
\multirow{2}{*}{\textbf{Base Model}} & \multicolumn{5}{c|}{\textbf{Standalone LLM (S3)}} & \multicolumn{5}{c|}{\textbf{MCP-Guard (S1-S3)}} & \multicolumn{2}{c}{\textbf{Improvement}} \\
& Acc & Prec & Rec & F1 & Time & Acc & Prec & Rec & F1 & Time & $\Delta$ F1 & Speedup \\
\midrule
\texttt{GPT-4o-mini} & \textbf{95.4} & 92.1 & \textbf{97.0} & \textbf{94.5} & 788.4 & 96.0 & 91.5 & 99.5 & \textbf{95.4} & 505.9 & +0.9 & 1.56$\times$ \\
\texttt{Deepseek-chat} & 92.3 & 88.9 & 92.8 & 90.8 & 3358.0 & 93.9 & 87.3 & \textbf{99.8} & 93.1 & 1988.2 & +2.3 & 1.69$\times$ \\
\texttt{Mistral:7B} & 82.8 & 87.9 & 67.4 & 76.3 & 435.4 & 90.8 & 83.3 & 97.0 & 89.6 & 157.3 & +13.3 & 2.77$\times$ \\
\texttt{Qwen2.5:0.5B} & 79.0 & 70.1 & 85.2 & 76.9 & \textbf{157.9} & 93.6 & 87.2 & 99.1 & 92.7 & 143.7 & +15.8 & 1.10$\times$ \\
\texttt{Llama3:8B} & 75.4 & \textbf{97.8} & 41.0 & 57.8 & 167.6 & \textbf{96.1} & \textbf{92.2} & 98.8 & \textbf{95.4} & \textbf{91.5} & +37.6 & 1.83$\times$ \\
\texttt{Tinyllama:1.1B} & 60.8 & 59.6 & 13.7 & 22.2 & 628.8 & 84.1 & 73.0 & 97.2 & 83.4 & 333.2 & \textbf{+61.2} & 1.89$\times$ \\
\texttt{Llama2:13B} & 43.6 & 39.9 & 73.6 & 51.7 & 1490.2 & 74.7 & 62.1 & 98.4 & 76.2 & 232.5 & +24.5 & \textbf{6.41$\times$} \\
\texttt{Gemma:7B} & 39.9 & 39.8 & 90.7 & 55.3 & 413.7 & 87.7 & 77.8 & 97.9 & 86.7 & 194.7 & +31.4 & 2.12$\times$ \\
\midrule
\rowcolor{cyan!10} \textbf{Average} & 71.1 & 72.0 & 70.2 & 65.7 & 930.0 & 89.6 & 81.8 & 98.5 & 89.1 & 455.9 & +23.4 & 2.04$\times$ \\
\bottomrule
\end{tabular}
\end{table*}

\end{document}